\documentclass[noreferee]{aa}
\usepackage{graphicx}
\usepackage{txfonts}
\usepackage{natbib}
\usepackage{graphicx}

\bibpunct{(}{)}{;}{a}{}{,} % to follow the A&A style

\newcommand{\boma}[1]{\mbox{\boldmath$#1$\unboldmath}}

\begin{document}

\title{Properties of the short period CoRoT-planet population II:\\
The impact of loss processes on planet masses\\
from Neptunes to Jupiters}

   \author{H. Lammer
          \inst{1}
          \and
          T. Penz
          \inst{2}
          \and
          G. Wuchterl
          \inst{3}
          \and
          H. I. M. Lichtenegger
          \inst{1}
          \and
          M. L. Khodachenko
          \inst{1}
          \and\\
          Yu. N. Kulikov
          \inst{4}
          \and
          G. Micela
          \inst{2}}

   %\offprints{H. Lammer,\\
   %\email{helmut.lammer@oeaw.ac.at}}

   \institute{Space Research Institute, Austrian Academy of Sciences, Schmiedlstrasse 6, A--8042 Graz, Austria\\
                \email{helmut.lammer@oeaw.ac.at, herbert.lichtenegger@oeaw.ac.at, maxim.khodachenko@oeaw.ac.at}
         \and
                INAF - Osservatorio Astronomico, Piazza del Parlamento 1, I--90134 Palermo, Italy\\
                \email{tpenz@astropa.inaf.it, giusi@astropa.unipa.it}
         \and
                 Th\"{u}ringer Landessternwarte Tautenburg, Sternwarte 5 D--07778 Tautenburg, Germany\\
                \email{gwuchterl@tls-tautenburg.de}
         \and
               Polar Geophysical Institute, Russian Academy of Sciences, Khalturina 15, 183010 Murmansk, Russian Federation\\
                \email{kulikov@pgi.ru}}

   %\date{Received December, 2006/ accepted ?}

\abstract{The orbital distance at which close-in exoplanets maintain
their initial mass is investigated by modelling the maximum expected
thermal and nonthermal mass loss rates over several Gyr. Depending
on an exosphere formation time and the evolution of the stellar
X-ray and EUV flux we expect that thermal evaporation at orbital
distances $< 0.05$ AU may be an efficient loss process for
hydrogen-rich exoplanets with masses $< 0.25M_{\rm Jup}$. Our
results indicate that nonthermal mass loss induced by Coronal Mass
Ejections of the host star can significantly erode weakly magnetized
short periodic gas giants. The observed exoplanets Gliese 876d at
0.0208 AU with a mass of $\sim 0.033M_{\rm Jup}$ and 55 Cnc e at
0.045 AU with a mass of $\sim 0.038M_{\rm Jup}$ could be strongly
eroded gas giants, while HD69830b, at 0.078 AU, HD160691d at 0.09 AU
and HD69830c at 0.18 AU belonged most likely since their origin to
the Neptune-mass domain. The consequences for the planetary
population predicted in paper I (Wuchterl et al. 2006) for CoRoTs
first field are: (1) for orbital distances $< 0.05$ AU (orbital
periods $<4$ days) weakly magnetized or highly irradiated gas giants
may loose a large fraction of their initial mass and completely
loose their gas envelopes. (2) Observed planetary mass spectra at
these periods that resemble the initial ones would indicate a major
effect of magnetic field protection and so far unknown thermospheric
cooling processes. (3) At distances $> 0.05 AU$ the impact of loss
processes is minor and the observed mass spectra should be close to
the initial ones.

}

\maketitle

\section{Introduction}
The upper atmospheres of short periodic exoplanets are affected strongly by the
stellar X-ray and EUV (XUV) radiation and plasma environment
\citep{Schneider1998}. Therefore, these bodies should experience high thermal
\citep{Lammer2003,Lecavelier2004,Yelle2004,
Vidal-Madjar2003,Baraffe2004,Tian2005,Erkaev2006,Penz2006} and nonthermal
\citep{Erkaev2005,Khodachenko2006} mass loss. Khodachenko et al (2006) studied
the minimum and maximum possible atmospheric erosion of the ``Hot Jupiter''
HD209458b due to Coronal Mass Ejections (CMEs) and found that this exoplanet,
which orbits a solar-like star at 0.046 AU could have been eroded to its
core-mass if its atmosphere were not protected by a strong magnetic field.
Because the mass loss depends on the strength of planetary magnetic fields,
which are estimated for slow rotating tidally locked ``Hot Jupiters'' to be in
a range of 0.005 -- 0.1 $\mathcal{M}_{\rm Jup}$ \citep{Griessmeier2004}, weakly
magnetized short periodic gas giants can experience large nonthermal mass loss
rates during their whole life time. The aim of this paper is to determine at
which distances exoplanets can maintain their initial mass, and where close
orbit gas giants can experience huge mass loss rates that may influence the
short period CoRoT-planet population.

\section{Exosphere-formation and the onset of planetary mass loss}
The overall evolution of a planet can be split into (1), the formation period
during which the planet is formed and its mass grows, and (2), the mature
period with the planet detached from gas reservoirs and exposed to loss
processes driven by its host star. During the formation period the planet is in
contact with a mass reservoir having certain pressure: the protoplanetary
nebula. During an isolated period planets are surrounded by vacuum with no
continuum pressure and an \emph{exosphere} as the interface. The overall mass
history of a planet is thus determined by pressure gradients as long as there
is a nebula pressure and by exosphere properties once there is no nebula
pressure.

The \emph{formation of the exosphere} separates these two eras and the regimes
of early mass-gain from the nebula and the eon long period of particle loss
processes. An exosphere exists if there is a layer in an atmosphere or
generally in a gas sphere where the mean free path of the particles is
sufficiently large to allow them to escape from a planet. Protoplanets fill
their Roche-lobe because that is where they are in contact with the nebula. For
a small planet/star mass-ratio the size of the Roche lobe can be well
approximated by the Hill-radius, $R_{\rm Hill} = d (M_{\rm pl}/[3 M_{\rm
star}])^{1/3}$, where $d$ is the orbital distance. Equating this to the mean
free path for hydrogen and assuming $P = n kT$ we obtain an exosphere formation
pressure $P_{\rm exo-form}$ at the Hill-sphere boundary
\begin{equation}
     \label{exo_formation_pressure}
     P_{\rm exo-form} = \frac{k T} {d \sigma_{\rm H}}
                             \left(\frac{3 M_{\rm star}}{M_{\rm pl}}
                             \right)^\frac{1}{3},
\end{equation}
where $\sigma_{\rm H}$ is the hydrogen collision cross section. Assuming an
exponential decay of the nebula pressure, $P_0$ with a time-scale $\tau_{\rm
neb}$, we calculate the time it takes for the lowest pressure around the planet
to reach $P_{\rm exo-form}$ and an exosphere to form within the Hill-sphere
\begin{equation}
t_{\rm exo-form} = \tau_{\rm neb} \ln \frac{P_0}{P_{\rm exo-form}}.
\end{equation}
We take $\sigma_{\rm H} = \pi a_0^2$, with the Bohr-radius, $a_0$ as
a lower limit for collisions under typical nebula-conditions. For
typical nebula pressures and temperatures at the planet formation
time, taken from the predicted CoRoT-planet population
\citep{Wuchterl2006}, we arrive at $t_{\rm exo-form} \approx 30 - \;
\tau_{\rm neb}$. With a typical global $\tau_{\rm neb}$ of $\sim 10$
Myr representing a compromise between various empirical methods
($\tau_{\rm neb}\sim 1$ -- 100 Myr) (see Hillenbrand et al. 2005 for
a recent update focused on dust IR-emission), one obtains average
exosphere formation-times, $t_{\rm exo-form} \approx 50$--300 Myr.
Thus exospheres may form at significant ages after the planets' host
star arrived at the Zero-Age-Main Sequence (ZAMS).

\section{Thermal mass loss}
In order to study the evolution of the maximum possible thermal
hydrogen loss rates from close-in Exosolar Giant Planets in a mass
range of $10^{29}$ g (EGP I) to $10^{30}$ g (EGP II) at different
orbital distances we estimate the thermal mass loss from the
evolution of the XUV flux which arrives from the host star at the
planet's orbit. We use in the present study a scaling law derived
from XUV flux observations of solar-like G stars with different age
$\Phi_{\rm XUV}=6.13 t^{-1.19} f_{\rm XUV}$ \citep{Ribas2005}, where
$t$ is the age of the star in Gyr, while $f_{\rm XUV}=8.5\times
10^{-4}$ W m$^{-2}$ is the flux at 1 AU averaged over the planetary
sphere. It is scaled to the different orbital distances by using an
$r^{-2}$ dependency. The maximum possible thermal mass loss rate
$\Gamma_{\rm th}$ can be calculated by assuming that an atmosphere
contains no efficient IR-cooling molecules and by using the
energy-limited equation which was originally derived for Earth-like
planets by Watson et al. (1981). Recently, Erkaev et al. (2006)
modified the energy-limited equation for the application to ``Hot
Jupiters''
\begin{equation}
\Gamma_{\rm th}=\frac{4 \pi R_{\rm pl} r_{\rm XUV}^2 \Phi_{XUV}}{ m
M_{\rm pl} G K},
\end{equation}
where $R_{\rm pl}$ and $M_{\rm pl}$ are the planetary radius and
mass, $K$ is a factor which is related to the Roche-lobe effect
\citep{Erkaev2006} and is $\approx 0.5$ for the considered planet,
$r_{\rm XUV}$ is the distance in the thermosphere where the optical
thickness $\tau_{\rm XUV} \rightarrow 1$ and the main part of the
XUV radiation is absorbed. One should note that $r_{\rm XUV}$ is
much closer to the planetary radius $R_{\rm pl}$ than the distance
used by Watson (1981). By using Eq. (3) with $r_{\rm XUV}$ of $\sim
1.3R_{pl}$ \citep{Yelle2004} we can calculate the maximum expected
thermal mass loss rate for HD209458b ($R_{\rm pl}=1.43$ $R_{\rm
Jup}$, $M_{\rm pl}=0.69$ $M_{\rm Jup}$) at present time. It is $\sim
1.7\times 10^{11}$ g s$^{-1}$ and appears to be in good agreement
with Vidal-Madjar et al. (2003), Yelle (2004) and Tian et al.
(2005). The mass loss rate for HD209458b at about $200$ Myr after
its host star arrived at the ZAMS is $\sim 6.0\times 10^{12}$ g
s$^{-1}$. We note that Lammer et al. (2003) applied Watson's
assumption for ``Hot Jupiters'' and overestimated the energy-limited
loss rate by an order of magnitude. One can see from Table 1 that
thermal evaporation may be an efficient loss process for close-in
gas giants at $d < 0.05$ AU with $M_{\rm pl}< 5\times 10^{29}$ g ($<
0.25M_{\rm Jup}$ and exosphere formation times $\leq 200$ Myr.
\begin{table}[t!]
\begin{center}
\begin{tabular}{c|c|c|c}
d [AU] & P [d] & $EGP I: L_{\rm th}$ [\%] & EGP II: $L_{\rm th}$ [\%] \\
\hline \hline
0.02  & 1   &  $\sim 87$  &  $\sim 8.7$  \\
0.05  & 4   &  $\sim 14$  &  $\sim 1.4$  \\
0.013 & 16  &  $\sim 2$   &  $\sim 0.2$  \\
\end{tabular}
\end{center}
\caption{Thermal mass loss in \% of the initial planetary mass of
EGP I and EGP II integrated over the history of the stellar system
for a representative exosphere formation time $t_{\rm
exo-form}\approx 200$ Myr.}
\end{table}
If one proceeds to longer exosphere formation times the integrated
mass loss is gradually decreasing. At larger orbital distances, the
thermal mass loss is $\leq 1$ \%. Thermal evaporation from close-in
lower mass exoplanets may be strong enough to remove their hydrogen
envelopes from their cores. One should also note that exoplanets
with orbits $< 0.02$ AU will experience an even higher thermal mass
loss due to the higher XUV flux and the mass loss enhancement due to
the Roche-lobe effect \citep{Erkaev2006}.

\section{CME-induced nonthermal mass loss}
Taking into account that tidal-locking of short periodic exoplanets
may result in weaker planetary magnetic moments, as compared to fast
rotating Jupiter-class planets at larger orbital distances
(Grie{\ss}meier et al. 2004), Khodachenko et al. (2006) found that
the encountering CME plasma may compress the magnetosphere and force
the magnetospheric standoff distance down to heights where
ionization and ion pick-up of the planetary neutral atmosphere by
the CMEs plasma flow takes place. Assuming for the G-type host star
of HD209458b the average CME occurrence rate as the one observed on
the Sun, Khodachenko et al. (2006) found that, depending on magnetic
protection ``Hot Jupiters'' at 0.045 AU could have lost over their
lifetime a mass from 2 \% up to more than $M_{\rm Jup}$. For
estimating the orbital distance at which Jupiter-class exoplanets
maintain the main part of their initial mass, we calculate the
``maximum'' possible CME-induced mass loss between orbital distances
of 0.015 -- 0.2 AU.

In the case of our Sun the maximum expected CME plasma density $n_{\rm CME}$ at
orbital distances $\leq 0.2$ AU is estimated from the analysis of CME
associated brightness enhancements in the white-light coronagraph images. By
using an analogy between the Sun and solar-type stars the maximum CME density
dependence on the orbital distance $d$ from the star can be assumed as a
power-law (e.g., Khodachenko 2006, and references therein) $n_{\rm CME}(d)= n_0
(d/d_0)^{-3.6}$, which for $n_0 = 5 \times 10^5 ... 5 \times 10^6 \mbox{
cm$^{-3}$}$ and $d_0 = 3 R_{\rm Sun}$ gives a good approximation for the values
estimated from the SOHO/LASCO coronograph images. The average mass of CMEs,
$M_{\rm CME}$, is $\approx 10^{15}$ g and the average duration of CMEs,
$\tau_{CME}$, at distances $(6...10) R_{\rm Sun}$ is $\approx 8$ h. The
collision rate between CMEs and ``Hot Jupiters'' can be estimated by
\cite{Khodachenko2006}
\begin{equation}
f_{\rm col} = \frac{\left(\Delta_{\rm CME} + \delta_{\rm pl}\right)
\sin\left[(\Delta_{\rm CME}+\delta_{\rm pl})/2\right]}{2\pi
\sin\Theta}f_{\rm CME},
\end{equation}
where $f_{\rm CME}$ is the CME occurrence rate, $\Delta_{\rm CME}$ and
$\delta_{\rm pl}$ are the CME and planetary angular sizes, $\Theta$ is the
latitude distribution of CME producing regions. Assuming an average stellar CME
occurrence rate $f_{\rm CME}$ of $\sim 3$ CMEs per day (Khodacheno et al. 2006;
and references therein), and taking the size, duration and latitude
distribution of CMEs close to the solar values ($\Delta_{\rm CME}=\pi/4$ to
$\pi/3$, $t_{\rm CME}=8$ h, $\Theta=\pi/3$), with the angular size $\delta_{\rm
pl}$ related to the orbital distance between 0.015--0.2 AU of 0.045 -- 0.0034
rad, one obtains a CME collision rate $f_{\rm col}$ of $\sim 0.17$ -- 0.3 hits
per day. Using these values an average total CME exposure time $t_{\rm
CME}\approx 1.2\times 10^{16}$ s during a period of 5 Gyr is obtained.

To calculate atmospheric erosion rates we apply the same time-dependent
numerical algorithm, which is able to solve the system of hydrodynamic
equations through the transonic point for calculation of the upper atmospheric
hydrogen density as used by Khodachenko et al. (2006) and Penz et al. (2006).
The obtained density profiles are in agreement with the density profiles
calculated by Yelle (2004) and Tian et al. (2005). For studying the expected
maximum possible atmospheric erosion due to CMEs we assume magnetospheres which
build obstacles against the CME plasma flow at substellar distances of 1.3, 1.5
and 2.0 $R_{\rm pl}$. For the CME-induced H$^+$ pick-up loss rates caused by
the CME plasma flow we take into account the ionization processes produced by
CMEs during the interaction with the neutral atmosphere above the assumed
planetary obstacles. The H$^+$ pick up loss calculations are performed with a
numerical test particle model which was successfully used for the simulation of
ion pick-up loss rates from various planetary atmospheres
\citep{Lichtenegger1995, Erkaev2005, Khodachenko2006}.
\begin{table}[t!]
\begin{center}
\begin{tabular}{c|c|c|c}
d [AU] & P [d] & Obstacle $\times R_{\rm pl}$ & $\Gamma_{\rm CME}$ [g s$^{-1}$] \\
\hline \hline
0.02 &  1.0 & 1.5        & $2.8\times 10^{14}$         \\
0.02 &  1.0 & 2.0        & $3.8\times 10^{13}$         \\ \hline
0.05 &  4.0 & 1.3        & $10^{14}$         \\
0.05 &  4.0 & 1.5        & $2.0\times 10^{13}$         \\
0.05 &  4.0 & 2.0        & $2.7\times 10^{12}$         \\ \hline
0.013 &  16.0 & 1.3        & $\times 10^{13}$         \\
0.013 &  16.0 & 1.5        & $1.5\times 10^{12}$         \\
0.013 &  16.0 & 2.0        & $3\times 10^{11}$         \\
\end{tabular}
\end{center}
\caption{Maximum CME-induced H$^+$ pick up ion mass loss rates per
CME collision for ``Hot Jupiters'' with different substellar
planetary magnetopause obstacles as a function of orbital distance.}
\end{table}
The calculation of the particle fluxes is performed by dividing the space above
and around the planetary obstacle onto a number of volume elements $\Delta
V_{\rm i}$. Production rates of planetary H$^+$ ions are obtained by
calculation of the CME plasma flow absorption along streamlines due to charge
exchange, photo-ionization, and electron impact ionization with the extended
upper atmosphere. The CME plasma flux $\Phi_{\rm CME}$ in a volume element
$\Delta V_{\rm i}$ at a given position $\boma{r}_{\rm i}$ with respect to the
planetary center is determined by
\begin{equation}
\Phi_{\rm CME}(\boma{r}_{\rm i})=\Phi_{\rm CME}^{(0)}(\boma{r}_{\rm
i}) e^{-\int\limits_\infty^{s_{\rm i}}\sum_{\rm H} n_{\rm
H}\sigma_{\rm H}^i ds},
\end{equation}
\begin{figure}[t]
\includegraphics[width=.49\textwidth]{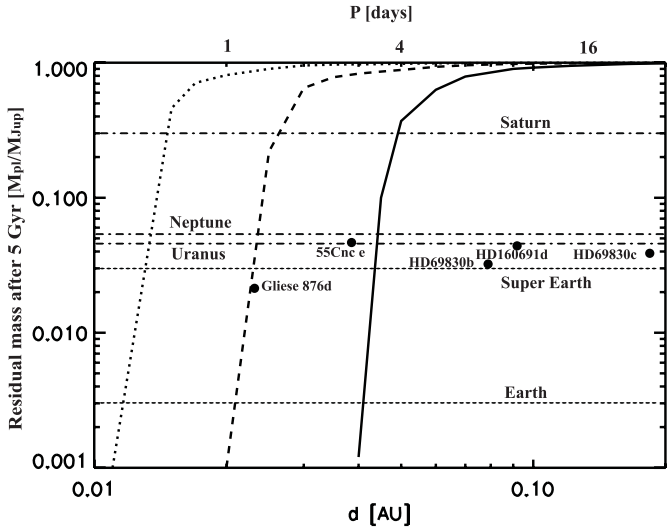}
\caption{Residual mass after 5 Gyr exposure of dense CMEs in units
of $M_{\rm Jup}$ of weakly magnetized ``Hot Jupiters'' as a function
of orbital distance compared with planetary obstacles at 1.3 $R_{\rm
pl}$ (solid line), 1.5 $R_{\rm pl}$ (dashed line), and 2.0 $R_{\rm
pl}$ (dotted line). The horizontal dashed-dotted lines indicate the
masses of Saturn, Uranus, Neptune, the horizontal dotted lines
corresponds to the mass of the Earth (lower line) and to the
``Super-Earth'' ($M_{\rm pl}\approx10M_{\rm Earth}$) mass domain
(upper line), and the black circles show the so far observed lower
mass close-in exoplanets.}
\end{figure}
where the integration is performed from the upstream CME plasma flow
to the corresponding point $s_{\rm i}$ at position $\boma{r}_{\rm
i}$ on the streamline. $\Phi_{\rm CME}^{(0)}$ is the unperturbed CME
plasma flux, $n_{\rm H}$ is the neutral hydrogen density as a
function of planetary distance, $\sigma_{\rm H}^i$ is the energy
dependent cross section of hydrogen ionization processes, and $ds$
is the line element along the streamline. The total planetary ion
production rate from atmospheric hydrogen atoms in each volume
element can be written as the sum $p^{\mbox{\tiny {\rm i}}}_{\rm
H^+}=p^{\mbox{\tiny {\rm ei}}}_{\rm H^+}+ p^{\mbox{\tiny {\rm
ce}}}_{\rm H^+}+p^{\mbox{\tiny $\gamma$}}_{\rm H^+}$, with
$p^{\mbox{\tiny {\rm ei}}}_{\rm H^+}$ the electron impact
ionization, $p^{\mbox{\tiny {\rm ce}}}_{\rm H^+}$ the charge
exchange rate, and $p^{\mbox{\tiny $\gamma$}}_{\rm H^+}$ the
photo-ionization rate. The total H$^+$ ion mass loss rate finally
becomes
\begin{equation}
\Gamma_{\rm CME}=\sum_i p^{\mbox{\tiny {\rm i}}}_{\rm H^+} \Delta
V_{\rm i} m_{\rm H^+}
\end{equation}
Table 2 shows the maximum expected CME-induced H$^+$ pick up mass loss rates
due to CME collisions from weakly magnetized ``Hot Jupiters'' with substellar
planetary obstacles at 1.3 $R_{\rm pl}$, 1.5 $R_{\rm pl}$, and 2.0 $R_{\rm pl}$
at various orbital distances. Fig. 1 shows the residual mass after 5 Gyr of the
maximum CME plasma exposure of weakly magnetized ``Hot Jupiters'' with
planetary obstacles at 1.3 $R_{\rm pl}$, 1.5 $R_{\rm pl}$, and 2.0 $R_{\rm pl}$
as a function of orbital distance compared to 5 known low-mass exoplanets
(Gliese 876d, 55 Cnc e, HD69830b, HD160691d, HD69830c)
(http://exoplanet.eu/catalog.php). The results obtained from the numerical ion
pick up test particle model simulations indicate that weakly magnetized gas
giants at orbital distances $\leq 0.05$ AU or periods $\leq 4$ days may lose a
mass equivalent to that of Jupiter during a 5 Gyr time period. By comparing the
results of the integrated CME-induced mass loss with the thermal evaporation
loss shown in Table 1, one can see that for magnetically unprotected planets
nonthermal loss processes are much more efficient than thermal evaporation.

Baraffe et al. (2004) compared the ratio of the mass loss timescale
$t_{\rm \dot{M}}$ to the thermal timescale $t_{\rm th}$, which is
characterized by the Kelvin-Helmotz timescale, found that, when
$t_{\rm \dot{M}}/t_{\rm th}$ becomes $< 1$ the evolution of the
``Hot Jupiter'' changes drastically which results in rapid expansion
of the planetary radius and enhances the mass loss. Depending on the
strength of the planetary magnetic field, XUV flux and exosphere
formation time, close-in gas giants at orbital distances $\leq 0.05$
AU may experience this violent mass loss effect.

One can also see from Fig. 1 that it is very unlikely that the three low mass
exoplanets  HD69830b, HD160691d, and HD69830c are remaining cores of eroded gas
giants. They could have lost their hydrogen-envelopes but their initial mass
was not much larger than that of Neptune. The two other low mass exoplanets
Gliese 876d and 55 Cnc e are located at orbital distances where they could have
been strongly affected by mass loss. Therefore, both exoplanets can be
remaining cores of eroded weakly magnetized gas giants.

\section{Conclusions}
Our study indicates that mass loss of weakly magnetized short
periodic ``Hot Jupiters'' can produce the observed masses of Gliese
876d and 55 Cnc e, while the three other known lower mass close-in
exoplanets HD69830b, HD160691d and HD69830c belonged most likely
since their origin to the Neptune-mass domain. The results of our
study indicate that only a combination of the size detection with
CoRoT and ground-based follow-up mass determinations together with
theoretical mass loss studies over the exoplanets' history can bring
reliable information on the statistics of remaining cores, shrinked
gas giants and ``unaffected'' lower mass exoplanets, like Super
Earths or Ocean planets. Furthermore, thermal evaporation from ``Hot
Jupiters'' at orbits $< 0.02$ AU, together with CME-triggered
nonthermal mass loss processes and tidal interactions
\citep{Patzold2002} can be a reason that so far no ``Hot Jupiter''
is observed at these close distances to its host star.

\begin{acknowledgements}
H. Lammer and Yu. N. Kulikov thank the AAS ``Verwaltungsstelle f\"ur
Auslandsbeziehungen'' and the RAS. T. Penz and G. Micela
acknowledges support by the Marie Curie Fellowship Contract No.
MTKD-CT-2004-002769 of the project ``The influence of stellar high
radiation on planetary atmospheres''. The authors also thank the
Austrian Ministry bm:bwk and ASA for funding the CoRoT project.
\end{acknowledgements}

\end{document}